# Amplitude dependence of image quality in atomically-resolved bimodal atomic microscopy

Hiroaki Ooe,[1,2 a)] Dominik Kirpal,[1] Daniel S. Wastl,[1] Alfred J. Weymouth,[1] Toyoko Arai,[2] and Franz J. Giessibl[1].

[1]*Institute of Experimental and Applied Physics, University of Regensburg D-93053 Regensburg, Germany*

[2]*Natural Science and Technology, Kanazawa University, Kanazawa, 920-1192 Ishikawa, Japan*

[a] electronic mail: hiroakiooe@se.kanazawa-u.ac.jp

In bimodal FM-AFM, two flexural modes are excited simultaneously. The total vertical oscillation deflection range of the tip is the sum of the peak-to-peak amplitudes of both flexural modes (sum amplitude). We show atomically resolved images of KBr(100) in ambient conditions in bimodal AFM that display a strong correlation between image quality and sum amplitude. When the sum amplitude becomes larger than about 200 pm, the signal-to-noise ratio (SNR) is drastically decreased. We propose this is caused by the temporary presence of one or more water layers in the tip-sample gap. These water layers screen the short range interaction and must be displaced with each oscillation cycle. Further decreasing the sum amplitude, however, causes a decrease in SNR. Therefore, the highest SNR in ambient conditions is achieved when the sum amplitude is slightly less than the thickness of the primary hydration layer.

Frequency modulation atomic force microscopy (FM-AFM)[1] is a powerful tool for investigating atomic-scale phenomena. The interaction between a tip at the end of an oscillating cantilever and a sample is measured via the frequency shift of the oscillation. If the oscillation amplitude is much larger than the decay length of the short range interaction, the tip spends little time within the short range interaction region, leading to a small contribution of the short range interaction to the detectable signal. It has been shown that in vacuum FM-AFM, the maximum signal-to-noise ratio (SNR) is achieved with an oscillation amplitude slightly larger than this decay length.[2] When investigating short range forces that decay at lengths comparable to interatomic distances, the highest SNR is achieved with amplitudes in the range from several tens to a few hundred picometers (small amplitudes).

Soft cantilevers that have a spring constant of $k < 100$ N/m (which is typical for commercial silicon cantilevers) require large amplitudes to prevent the tip from crashing into the surface at close distance (so-called "jump-to-contact").[3] For this reason, atomic resolution measurements with soft cantilevers require the use of large amplitudes from one nanometer to tens of nanometers.[4-6] One way to achieve controllable small amplitudes with soft cantilevers is to use a higher flexural mode which provides a much higher effective stiffness than the fundamental mode.[7] Theoretically, the effective stiffness of the second flexural mode is about 40 times higher than in the first flexural mode and the resonance frequency is about 6.2 times higher.[8] This can be implemented with bimodal AFM,[9,10] in which the first flexural mode is excited at a large amplitude and the second flexural mode at a small amplitude to detect short range interactions.

Bimodal AFM has been very successful in ambient and vacuum environments.[10-12] In ambient environments or liquids, the cantilever must oscillate through the liquid or through water condensation layers.[12-15] It was shown that the small oscillation of the second flexural mode could be used to increase sensitivity to materials properties.[16,17] Several groups have applied this technique to biological samples, including antibodies[12] and proteines.[10] Schwenk and coworkers used bimodal AFM to increase MFM contrast of magnetic samples.[20,21] Kawai and coworkers explicitly demonstrated the advantage of a higher flexural mode oscillating at smaller amplitudes (amplitudes less than 100 pm) with a standard Si cantilever on a KBr(100) surface in UHV.[22] Moreno and coworkers used this to achieve intramolecular resolution in UHV conditions at low temperature.[11] More recently, Santos and coworkers have started to consider the advantages of small oscillations in both flexural modes.[17]

It is expected for atomic resolution that the ideal bimodal measurement should be acquired with small amplitudes in both the first and second flexural modes.[17] This requires the use of a much stiffer sensor.

In this Letter, we present data acquired with a qPlus sensor with a spring constant of $k = 1800$ N/m.[24] This high stiffness allows oscillation amplitudes of the first flexural mode smaller than one angstrom.[14-15] We performed measurements in ambient condition on KBr(100). The amplitude of the first flexural mode, $A_1$ and of the second flexural mode, $A_2$, were independently set. These amplitudes were calibrated with a thermal spectrum and the ratio of the deflection sensitivity of two flexural modes.[8,24-26] The frequency shifts of the first flexural mode, $\Delta f_1$, and of the second flexural mode, $\Delta f_2$, were recorded in quasi-constant height mode using low gain integral feedback to compensate for thermal drift. We used a sensor with a resonance frequency of the first flexural mode of $f_1 = 32596.7$ Hz and quality factor $Q_1 = 2906$. The second flexural mode had a resonance frequency of $f_2 = 194858.2$ Hz and a quality factor of $Q_2 = 1848$.

In ambient conditions, water condenses on all surfaces. Near the surface, it forms ordered hydration layers with a thickness of $\sim 200$ - $310$ pm.[13-15,29-33] In previous work, the ideal amplitude of oscillation was determined for single-mode FM-AFM measurements in ambient conditions.[13-15] On the KBr(100) surface, the highest SNR was observed with an amplitude of $A \sim 75$pm.[14] With smaller amplitudes, the signal becomes noisier due to instrumental noise.[1,26,27] With larger amplitudes, the SNR suffers for two reasons: The average tip-sample distance becomes larger, reducing the signal, and water molecules come between the tip and sample. The tip then needs to penetrate the hydration layer during each oscillation cycle and the water molecules screen the short range interaction.[14] Because of these effects, the SNR is enhanced when the peak-to-peak amplitude is slightly smaller than the thickness of a single hydration layer.[15]

Figure 1 shows single-mode FM-AFM and bimodal FM-AFM measurements. The oscillation models of the first and second flexural modes are shown in Figure 1 (a) and (b). First we collected single-mode images, exciting either the first or the second flexural mode. Figure 1(c) is a $\Delta f_1$ image taken with only the first flexural mode excited at $A_1 = 75$ pm and Figure 1(d) is a $\Delta f_2$ image taken with only the second flexural mode excited at $A_2 = 75$ pm. Atomic resolution can clearly be seen in both images.

We then investigated if the two modes influence each other. To do this, we first acquired $\Delta f_1$ data with only the first flexural mode excited, then also excited the second mode. Figures 1 (e) and (f) show simultaneously acquired $\Delta f_1$ and $\Delta f_2$. The slow scan direction of these imaging was downward. Down to line A, only the first flexural mode was excited at $A_1 = 75$ pm, and atomic resolution can clearly be seen in $\Delta f_1$. From line A down, the second flexural mode is also excited at $A_2 = 75$ pm and the $\Delta f_2$ controller was turned on from line B. With both modes excited, the $\Delta f_1$ image becomes much weaker.

Next we acquired data with both first and second flexural modes excited at the same amplitude. Figure 2 shows images of $\Delta f_1$ and $\Delta f_2$ with both flexural modes excited at amplitudes of 75 pm, 53 pm and 40 pm. When $A_1 = A_2 = 75$ pm, the (a) $\Delta f_1$ image and (b) $\Delta f_2$ image show faint atomic contrast similar to that in Figure 1 (e). The images improve when the amplitudes are decreased, as can be seen in Figure 2 (c) and (d), for which $A_1 = A_2 = 53$ pm. Very clear images are obtained when $A_1 = A_2 = 40$ pm, shown in Figure 2 (e) and (f).

Similar to previous findings in vacuum,[2] we find an optimal SNR for amplitudes in the sub-Angstrom level. However, empirically we find a notable difference to the decrease of SNR when increasing the amplitude beyond its optimal value $A_{opt}$. In vacuum, SNR decreases quite shallow at a rate of approximately[2] $(A_{opt}/A)^{0.5}$. In ambient environments with a liquid adsorption layer, we find a much stronger decay of image quality with sum amplitude. The sum amplitude is the vertical range that is covered by the oscillating cantilever: $z_{p-p} = 2\ (A_1 + A_2)$. We propose that using sum amplitudes greater than half the thickness of the first hydration layer (approx. 200 pm) allows water molecules to penetrate the gap between the tip apex and the sample, reducing the image quality.

We then varied $A_1$ and $A_2$, keeping $z_{p\text{-}p}$ less than the thickness of a single hydration layer. In Figure 3(a) and (b), atomic contrast can be seen in both $\Delta f_1$ and $\Delta f_2$. $A_1 = 60$ pm $> A_2 = 15$ pm. Correspondingly, $\Delta f_1$ shows higher SNR. In Figure 3(c) and (d), $A_1 = A_2$ and the SNR of the two images are similar. Finally, in Figure (e) and (f), $A_1 = 15$ pm $< A_2 = 60$ pm, and the corresponding $\Delta f_2$ image has a higher SNR. These results show that SNR is higher with larger amplitudes. Instrumental noise decreases with increasing amplitude. Therefore, for each mode, larger amplitudes correspond to lower noise, as expected when considering the noise contributions in FM-AFM.[1,27,28]

In vacuum, the optimal amplitude is given by the balance of a more precise frequency measurement for larger amplitudes at the cost of a smaller frequency shift signal for larger amplitudes, resulting in an optimal amplitude that is approximately given by the decay length of the short-range interaction.[2] In ambient conditions, the noise in frequency measurement also decreases for larger amplitudes, but the frequency shift signal induced by short-range interactions drops rapidly once the sum amplitude is large enough to admit water molecules in the tip-sample gap. The result is that the ideal amplitudes for bimodal FM-AFM follow the same pattern as for single-mode FM-AFM measurements. In ambient conditions, the sum amplitude must be smaller than the thickness of a hydration layer to ensure that the tip does not leave and re-penetrate a hydration layer with each cycle. At the same time, the amplitude has to be as large as possible to reduce the noise. The resolution of each mode can be increased by increasing its amplitude up to the ideal sum amplitude.

In this study, we investigated the effect of the amplitude of the first and second flexural modes on the image quality in bimodal FM-AFM with small amplitudes in ambient conditions. Two orthogonal flexural modes can have a strong influence on each other. This is due to the hydration layer of sample surface. We showed that for this system, maximizing the SNR for both $\Delta f_1$ and $\Delta f_2$ results in the requirement that $A_1 = A_2$. Our results supporting that conventional bimodal AFM might also benefit from stiffer cantilevers that enable a smaller fundamental amplitude.

**Acknowledgements**

Funding was provided by Deutsche Forschungsgemeinschaft under GRK 1570 and by "Strategic Young Researcher Overseas Visits Program for Accelerating Brain Circulation" from the Japan Society for the Promotion of Science. The authors gratefully acknowledge the support for this study provided by Kanazawa University SAKIGAKE Project.

**References**
[1] T. R. Albrecht, P. Grütter, D. Horne, and D. Rugar, J. Appl. Phys. **69**, 668 (1991).
[2] F. J. Giessibl, H. Bielefeldt, S. Hembacher, and J. Mannhart, Appl. Surf. Sci. **140**, 352 (1999).
[3] F. J. Giessibl, Phys. Rev. B **56**, 16010 (1997),

[4] F. J. Giessibl, Science **267**, 68 (1995).

[5] S. Kitamura, and M. Iwatsuki, Jpn. J. Appl. Phys. **34**, L145 (1995).

[6] K. Fukui, H. Onishi, and Y. Iwasawa, Phys. Rev. Lett. **79**, 4202 (1997).

[7] S. Kawai, S. Kitamura, D. Kobayashi, S. Meguro, and H. Kawakatsu, Appl. Phys. Lett., **86**, 193107 (2005).

[8] S. Rast, C. Wattinger, U. Gysin, and E. Meyer, Rev. Sci. Instrum. **71**, 2772 (2000).

[9] T. R. Rodriguez, and R. Garcia, Appl. Phys. Lett. **84**, 449 (2004).

[10] R. Garcia, E. T. Herruzo, Nature Nanotechnology **7**, 217 (2012).

[11] C. Moreno, O. Stetsovych, T. K. Shimizu, and O. Custance, Nano Lett. **15**, 2257 (2015).

[12] N. F. Martinez, J. R. Lozano, E. T. Herruzo, F. Garcia, C. Richter, T. Sulzbach, and R. Garcia, Nanotechnology, **19**, 384011 (2008).

[13] D. S. Wastl, M. Judmann, A. J. Weymouth, and F. J. Giessibl, ACS Nano, **9**, 3858 (2015).

[14] D. S. Wastl, A. J. Weymouth, and F. J. Giessibl, Phys. Rev. B **87**, 245415 (2013).

[15] D. S. Wastl, A. J. Weymouth, and F. J. Giessibl ACS Nano, **8**, 5233 (2014)

[16] N. F. Martinez, S. Patil, J. R. Lozano, R. Garcia, Appl. Phys. Lett. **89**, 153115 (2006).

[17] S. Santos, Appl. Phys. Lett. **103**, 231603 (2013).

[18] S. D. Solares, and G. Chawla, J. Appl. Phys. **108**, 054901 (2010).

[19] G. Chawla, and S. D. Solares, Appl. Phys. Lett. **99**, 074103 (2011).

[20] J. Schwenk, M. Marioni, S. Romer, N. R. Joshi, and H. J. Hug, Appl. Phys. Lett. **104** 112412 (2014).

[21] J. Schwenk, X Zhao, M. Baconi, M. A. Manioni, S. Romer, and H. J. Hug, Appl. Phys. Lett. **107**, 132407 (2015)

[22] S. Kawai, T. Glatzel, S. Koch, B. Such, A. Baratoff, and E. Meyer, Phys. Rev. Lett. **103**, 220801 (2009).

[23] S. Santos, V. Barcons, H. K. Christenson, D. J. Billingsley, W. A. Bonass, J. Font, and N. H. Thomson, Appl. Phys. Lett. **103**, 063702 (2013).

[24] F. J. Giessibl, Appl. Phys. Lett. **76**, 1470 (2000).

[25] J. Welker, F. F. Elsner, and F. J. Giessibl, Appl. Phys. Lett. **99**, 084102 (2011).

[26] F. J. Giessibl, Rev. Mod. Phys. **75**, 949 (2003).

[27] K. Kobayashi, H. Yamada, and K. Matsushige, Rev. Sci. Instrum. **82**, 0337025 (2011).

[28] M. Luna, F. Rieutord, N. A. Melman, Q. Dai, and M. Salmeron, J. Phys. Chem. A **102**, 6793 (1998).

[29] S. Jeffery, P. M. Hoffmann, J. B. Pethica, C. Ramanujan, H. Ozgur Ozer, and A. Oral, Phys. Rev. B **70**, 054114 (2004).

[30] T. Fukuma, M. J. Higgins, and S. P. Jarvis, Biophys. J. **92**, 3603 (2007).

[31] K. Kimura, S. Ido, N. Oyabu, K. Kobayashi, and Y. Hirata, J. Chem. Phys. **132**, 194705 (2010).

[32] J. N. Israelachvili and R. M. Pashley, Nature (London) **306**, 249 (1983).

[33] J. I. Kilpatrick, S. Loh, and S. P. Jarvis, J. Am. Chem. Soc. **135**, 2628 (2013).

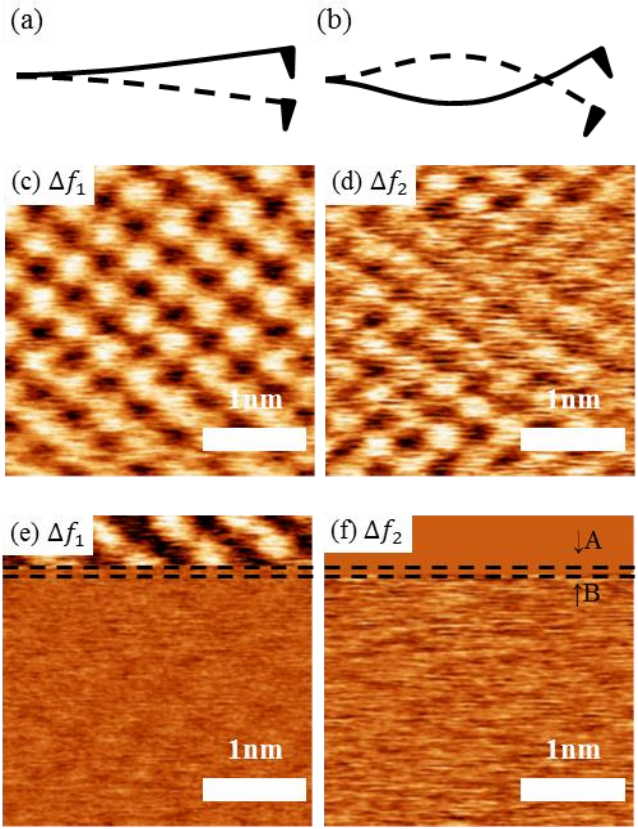

FIG. 1 Schematics of (a) first and (b) second flexural mode. (c) $\Delta f_1$ image with only the first flexural mode excited at $A_1 = 75$ pm. (d) $\Delta f_2$ with only the second flexural mode excited at $A_2 = 75$ pm. (e) $\Delta f_1$ and (f) $\Delta f_2$ images simultaneously acquired. Up to line A, only the first mode was excited at $A_1 = 75$ pm. Past line B, both modes were excited at $A_1 = A_2 = 75$ pm. For clarity, all images were line-flattened. The raw data with scale of frequency shift are shown in Figure 1 of supplementary material.

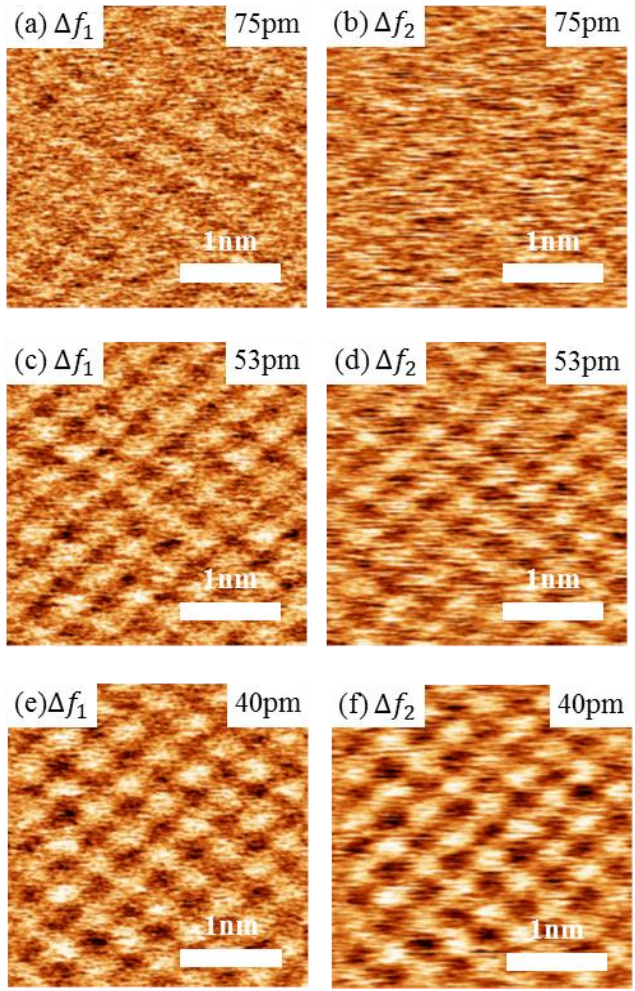

FIG. 2 Bimodal FM-AFM images taken in which $A_1 = A_2$. (a and b) $A_1 = A_2 = 75$ pm (c and d) $A_1 = A_2 = 53$ pm (e and f) $A_1 = A_2 = 40$ pm. Images are line-flattened for clarity. The raw data with scale of frequency shift are shown in Figure 2 of supplementary material.

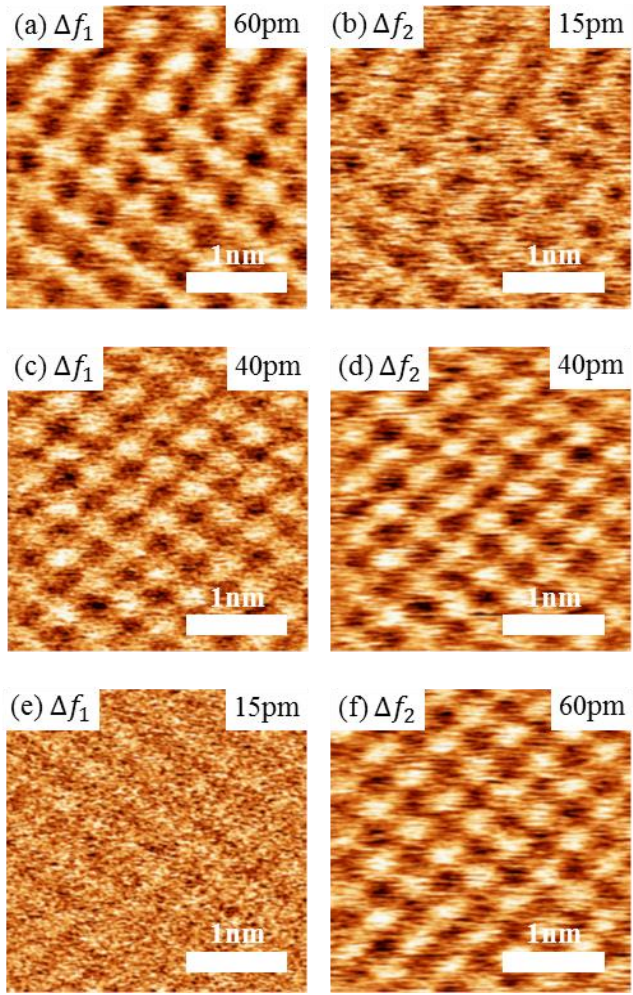

FIG.3 A survey of images taken with different $A_1$ and $A_2$ values in $A_1 + A_2 \sim 80$ pm. (a) $\Delta f_1$ image with $A_1$ = 60 pm, and (b) $\Delta f_2$ image with $A_2$ = 15 pm. (c) $\Delta f_1$ image with $A_1$ = 40 pm, and (d) $\Delta f_2$ image with $A_2$ = 40 pm. (e) $\Delta f_1$ image with $A_1$ = 15 pm, and (f) $\Delta f_2$ image with $A_2$ = 60 pm. All images were line flattened for clarity. The raw data with scale of frequency shift are shown in Figure 3 of supplementary material.

For Supplementary Material.

Figure 1 Raw data

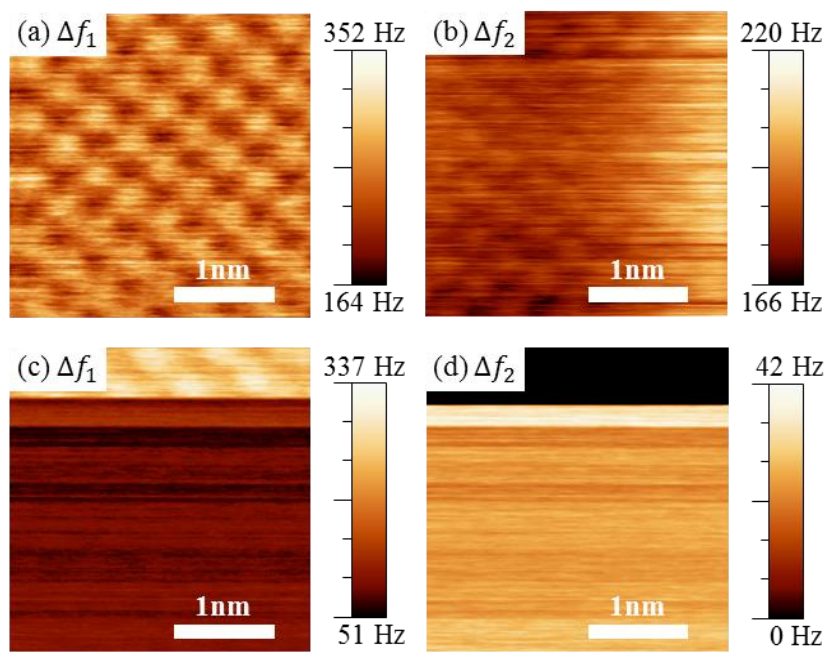

Figure 2 Raw data

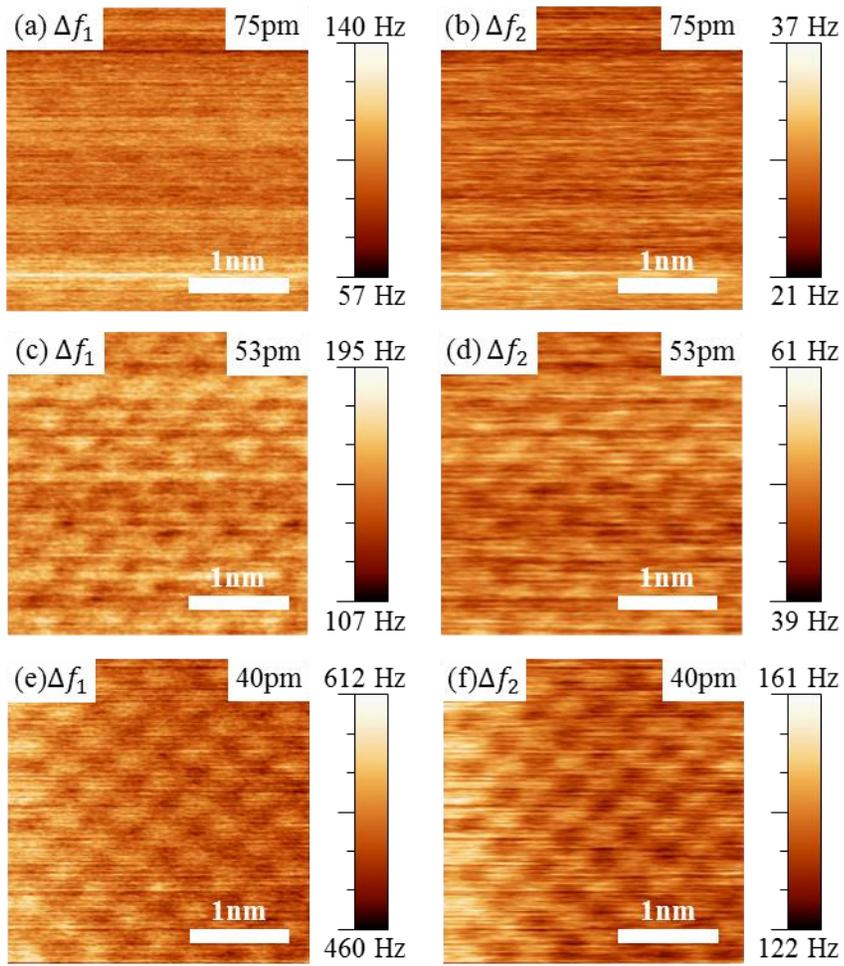

Figure 3 Raw data

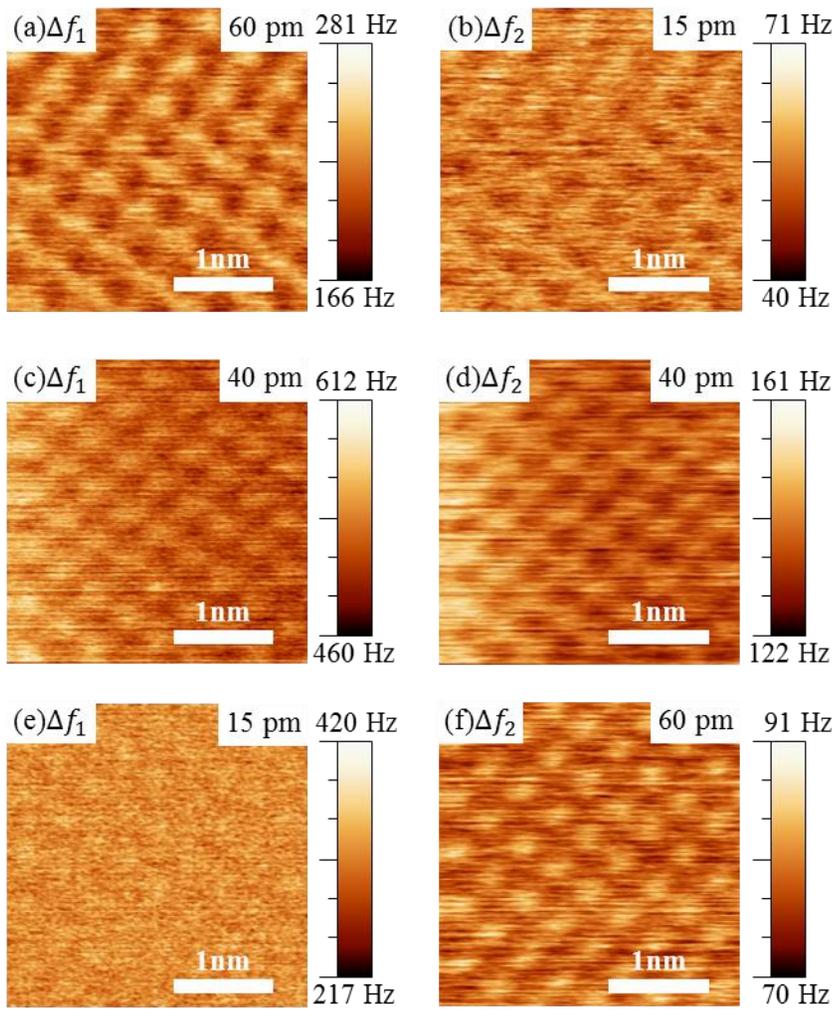